\documentclass[12pt]{article}

\usepackage{scicite}
\usepackage{amsmath}
\usepackage{times}

\usepackage{graphicx}
\usepackage{epstopdf}
\usepackage{braket}

\newenvironment{sciabstract}{
\begin{quote}
\bf}
{\end{quote}}

\title{High-fidelity laser-free universal control of two trapped ion qubits} 

\author{R. Srinivas$^{1,2,\dagger,\ast}$, S. C. Burd$^{1,2,\xi}$, H. M. Knaack$^{1,2}$, R. T. Sutherland$^{3,4}$, \\
A. Kwiatkowski$^{1, 2}$, S. Glancy$^{1}$, E. Knill$^{1,5}$, D. J. Wineland$^{1,2,6}$ \\
D. Leibfried$^{1}$, A. C. Wilson$^{1}$, D. T. C. Allcock$^{1,2,6}$, and D. H. Slichter$^{1,\ast}$.
\\
\\
\normalsize{$^{1}$National Institute of Standards and Technology, Boulder, CO 80305, USA}\\
\normalsize{$^{2}$Department of Physics, University of Colorado, Boulder, CO 80309, USA}\\
\normalsize{$^{3}$Physics Division, Physical and Life Sciences, Lawrence Livermore National Laboratory,}\\
\normalsize{Livermore, CA 94550, USA}\\
\normalsize{$^{4}$Department of Electrical and Computer Engineering, Department of Physics and Astronomy,}\\
\normalsize{University of Texas at San Antonio, San Antonio, TX 78249, USA}\\
\normalsize{$^{5}$Center for Theory of Quantum Matter, University of
Colorado, Boulder, Colorado 80309, USA}\\
\normalsize{$^{6}$Department of Physics, University of Oregon, Eugene, OR 97403, USA}\\
\\
\\
\normalsize{$^\ast$To whom correspondence should be addressed;}
\\ 
\normalsize{E-mail:  raghavendra.srinivas@colorado.edu; daniel.slichter@nist.gov.}
}

\date{}

\begin{document}

\baselineskip24pt


\maketitle


\begin{sciabstract}

Universal control of multiple qubits\textemdash the ability to entangle qubits and to perform arbitrary individual qubit operations\textemdash is a fundamental resource for quantum computation, simulation, and networking. Here, we implement a new laser-free scheme for universal control of trapped ion qubits based on microwave magnetic fields and radiofrequency magnetic field gradients. We demonstrate high-fidelity entanglement and individual control by creating symmetric and antisymmetric two-qubit maximally entangled states with fidelities in the intervals [0.9983,~1] and [0.9964,~0.9988], respectively, at 68\% confidence, corrected for state initialization error. This technique is robust against multiple sources of decoherence, usable with essentially any trapped ion species, and has the potential to perform simultaneous entangling operations on many pairs of ions without increasing control signal power or complexity.
\end{sciabstract}

Universal unitary control of quantum systems is essential for quantum computing~\cite{Jozsa1997}, simulation~\cite{Georgescu2014}, and networking~\cite{Kimble2008}. Universal control can be achieved through a combination of unitary entangling operations on pairs of qubits and single-qubit rotations on individual qubits~\cite{Barenco1995}. Trapped-ion entangling interactions typically rely on an effective qubit-qubit coupling mediated by the shared motion of the ions~\cite{Cirac1995, Milburn2000, Sorensen1999, Sorensen2000, Leibfried2003}. Realizing this effective coupling requires control fields at the ions with strong spatial gradients on the length scale of the ions' zero-point motion (usually a few nm), commonly generated using laser light with wavelengths of a few hundred nm.  Laser light can also be tightly focused to illuminate specific ions, providing individual qubit control~\cite{Naegerl1999}.  Laser-based universal control~\cite{SchmidtKaler2003, Debnath2016, Wright2019, Erhard2019} and two-qubit entanglement generation~\cite{Gaebler2016, Ballance2016} in trapped ion qubits have been demonstrated, with Refs.~\cite{Gaebler2016, Ballance2016} reporting fidelities for symmetric Bell states, corrected for errors not induced by the entangling operation, in the intervals of [0.9982,~0.9996] and [0.9988,~0.9996] at 68\% confidence, respectively, representing the current state of the art for any quantum system.  The achievements in Refs.~\cite{SchmidtKaler2003, Debnath2016, Wright2019, Erhard2019, Gaebler2016, Ballance2016} relied on specialized high-performance laser systems, and the record fidelities in Refs.~\cite{Gaebler2016, Ballance2016} were limited primarily by off-resonant photon scattering~\cite{Ozeri2007}. Laser-free trapped-ion entangling operations eliminate photon scattering errors and can harness radiofrequency and microwave technology developed for wireless communications. While laser-free entangling schemes have been proposed~\cite{Wineland1998, Mintert2001, Ospelkaus2008} and demonstrated using microwave-frequency~\cite{Ospelkaus2011,Harty2016,Hahn2019,Zarantonello2019} or static~\cite{Khromova2012, Weidt2016} magnetic field gradients, the highest-fidelity laser-free entangling interactions to date~\cite{Harty2016, Zarantonello2019} had approximately 2 to 3 times larger corrected Bell state infidelities than the laser-based interactions in Refs.~\cite{Ballance2016, Gaebler2016} and were more than an order of magnitude slower.

Here, we demonstrate a new laser-free entangling method using an oscillating near-field radiofrequency magnetic field gradient, achieving a symmetric Bell state fidelity in the interval of {[0.9983,\,1]} with 68\% confidence, corrected for initialization error. This fidelity is statistically indistinguishable from the highest-fidelity laser-based demonstrations, and the entangling operation is about four times faster than the previous highest-fidelity laser-free demonstrations. Our scheme is intrinsically robust to decoherence of both the ions' qubit and motional states~\cite{Sutherland2019, Hayes2012}, and enables the individual qubit addressing required for universal control using the same radiofrequency control fields that perform the entangling interaction.  We use this universal control to create antisymmetric Bell states\textemdash which requires individual qubit control\textemdash with fidelity in the interval [0.9964,~0.9988] with 68\% confidence, corrected for initialization error, which is, to the best of our knowledge, the highest to date in any qubit platform.

\begin{figure}[h]
\includegraphics[width=0.75\textwidth]{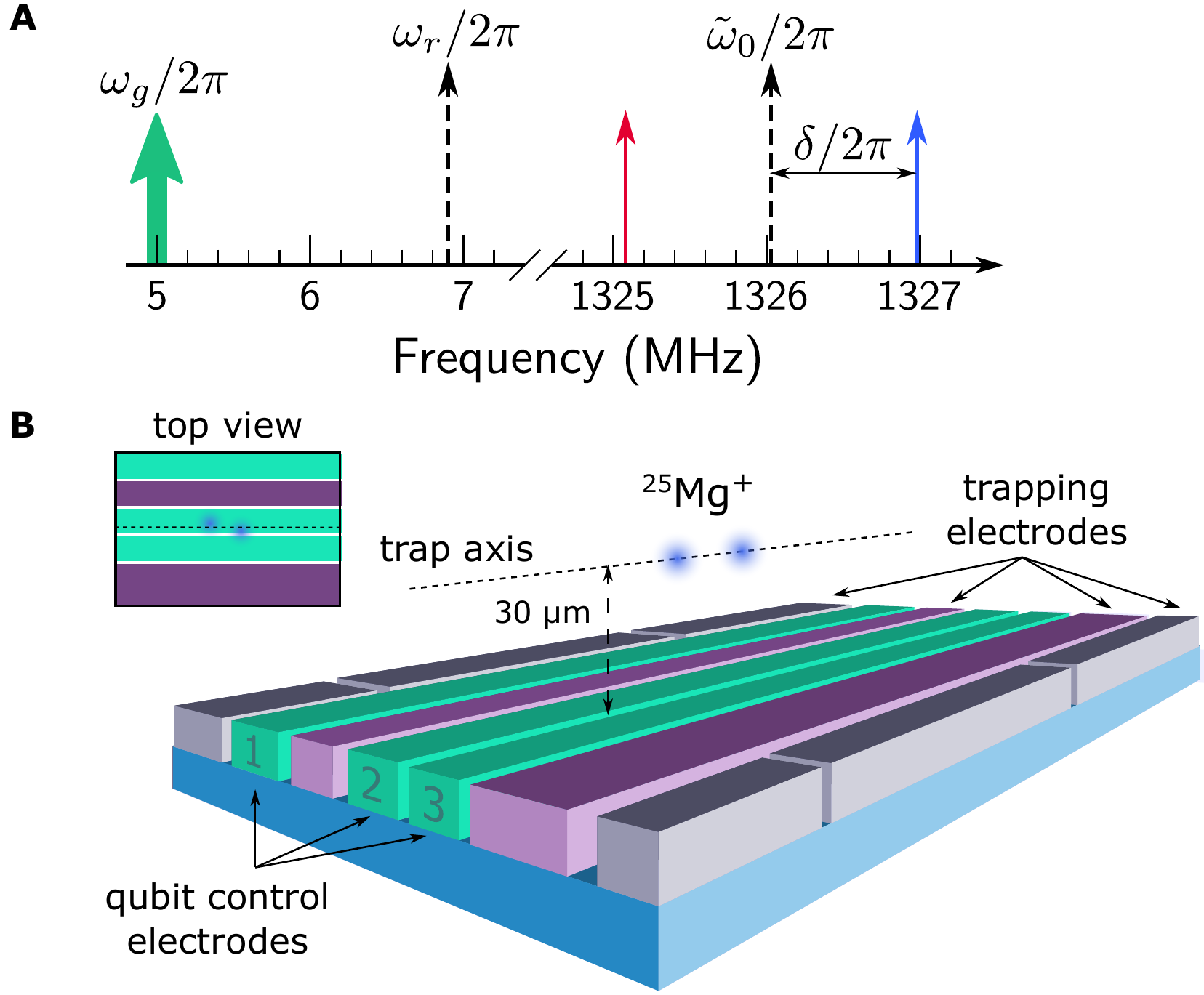}
\centering
\caption{\label{fig_general}
\textbf{(A)}  Spectrum of applied currents. We apply currents of approximately 1 A peak at ${\omega_g=2\pi\times5}$\,MHz, close to the ions' motional frequency $\omega_r\approx2\pi\times6.9$ MHz, to create a magnetic field gradient at the ions~\cite{supplementary}. We also apply two much weaker currents (tens of mA peak), symmetrically detuned by $\delta\approx(\omega_r-\omega_g)/2$ from the qubit frequency $\tilde{\omega}_0\approx2\pi\times1.326$\,GHz, to generate the entangling interaction. We indicate the motional mode and qubit frequencies with dashed arrows.
\textbf{(B)} 3D perspective view of the central region of the surface electrode ion trap (not to scale). For these experiments, we trap two $^{25}$Mg$^+$ ions approximately 30$\,\mu$m above the trap surface. The MHz and GHz currents used to generate the entangling interaction are driven along the qubit control electrodes (green, numbered $1-3$) and create magnetic fields and field gradients transverse to the trap axis, while the confining potential is created by oscillating and static voltages applied to the purple and gray trapping electrodes, respectively. We can apply static electric fields to rotate the ion crystal slightly with respect to the qubit control electrodes (inset), such that the two ions experience different ac Zeeman shifts from the magnetic field at $\omega_g$. The same ion crystal configuration can be used both for entanglement and individually addressed control.}
\end{figure}

The entangling operation relies on control signals at three frequencies, as shown in Fig.~\ref{fig_general}A.  A strong magnetic field gradient with amplitude 152(15) T/m oscillating at $\omega_g=2\pi\times5$\,MHz, close to the frequency $\omega_r$ of one of the ions' out-of-phase radial (transverse to the trap axis) motional modes at $\omega_r\approx 2\pi\times6.9$\,MHz, is combined with two additional weaker microwave magnetic fields, symmetrically detuned by $\delta$ from the qubit frequency ${\tilde{\omega}_0\approx2\pi\times1.326}$\,GHz, which is shifted from its nominal value of $\omega_0$ by residual magnetic fields oscillating at $\omega_g$. Previous laser-free entanglement demonstrations have required two high-power signals at GHz (rather than MHz) frequencies to generate large gradients~\cite{Ospelkaus2011, Harty2016, Zarantonello2019}, or eight microwave fields (four per qubit) along with a strong static magnetic field gradient~\cite{Weidt2016}. 

Choosing $\delta=(\omega_r-\omega_g)/2+\Delta/2$, where $|\Delta|\ll|\omega_r-\omega_g|$ is a small frequency offset, the slowly-rotating terms generate the interaction~\cite{Sutherland2019, supplementary}

\begin{align}
\label{eq_hamiltonian}
\hat{H}_I(t) = \hbar\Omega_g J_2\left(\frac{4\Omega_\mu}{\delta}\right)(
\hat{\sigma}_{z1}-\hat{\sigma}_{z2})(\hat{a}e^{i\Delta t}+\hat{a}^\dagger e^{-i\Delta t}),
\end{align}

\noindent which is used to implement a geometric phase gate~\cite{Milburn2000, Sorensen1999, Sorensen2000, Leibfried2003}, entangling the ion states via their shared motion with an effective $\hat{\sigma}_{z1}\hat{\sigma}_{z2}$ coupling. Here $\Omega_g$ and $\Omega_\mu$ are proportional to the amplitude of the radiofrequency gradient and the microwave fields, respectively~\cite{Srinivas2019, Sutherland2019}, $J_n$ is the $n^{th}$ Bessel function of the first kind, the Pauli operator $\hat{\sigma}_{zi}$ acts on ion $i$, and $\hat{a} (\hat{a}^\dagger)$ are the annihilation (creation) operators for the ions' selected motional mode.  By tuning the amplitude $\Omega_{\mu}$ of the two microwave fields to an appropriate value (an ``intrinsic dynamical decoupling'' or ``IDD'' point), the qubits are dynamically decoupled from dephasing noise at frequencies well below $\delta$, without requiring any additional control fields~\cite{Sutherland2019}. The microwave fields modulate the qubit state such that the effect of low-frequency dephasing noise on the qubit is multiplied by a prefactor of $J_0\left(4\Omega_\mu/\delta\right)$; the IDD points occur when $\Omega_
\mu$ is set such that $J_0\left(4\Omega_\mu/\delta\right)=0$.  Errors due to static qubit frequency offsets, which are proportional to $\hat{\sigma}_{zi}$ and thus commute with the entangling interaction, can also be mitigated by performing global qubit $\pi$ pulses~\cite{supplementary}. We can use these $\pi$ pulses to simultaneously implement Walsh modulation, which provides robustness to offsets and drifts in the motional frequency or in the control field amplitudes~\cite{Hayes2012}. This combination of techniques yields an entangling interaction with substantial protection against decoherence of both the qubit and the ion motion, as well as experimental miscalibrations.  

The experimental setup is similar to that in Ref.~\cite{Srinivas2019}, with radiofrequency and microwave control currents, as well as trapping voltages, applied to electroplated gold electrodes on a surface-electrode trap as shown in Fig.~\ref{fig_general}B. The trap is cooled to $\approx$~15\,K, and we perform our operations on two $^{25}$Mg$^+$ ions held approximately 30\,$\mu$m above the trap surface. We use the ${\ket{F=3,m_F=3}\equiv\ket{\downarrow}}$ and ${\ket{F=2, m_F=2}\equiv\ket{\uparrow}}$ states within the ions' $^2S_{1/2}$ ground state hyperfine manifolds as our qubit states.  We present complete details of the experimental setup in the supplementary material~\cite{supplementary}.  

\begin{figure}[t]
\includegraphics[width=0.9\textwidth]{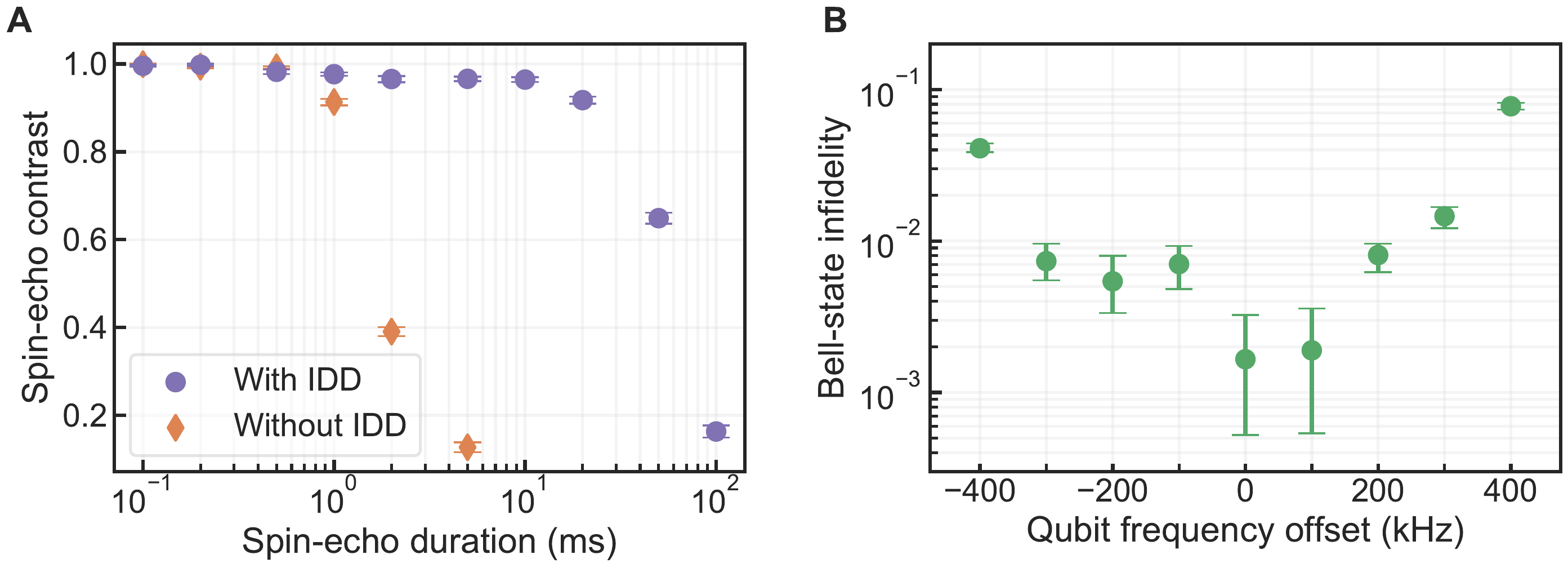}
\centering
\caption{\label{fig_idd}
\textbf{(A)} Single-ion qubit coherence with and without intrinsic dynamical decoupling (IDD). We plot spin-echo contrast as a function of the total duration of both spin-echo arms. Data are shown either with (purple circles) or without (orange diamonds) application of microwave fields tuned to the IDD point during the spin-echo arms. \textbf{(B)} Bell-state infidelity versus qubit frequency offset. The qubit frequency is offset from the midpoint between the two microwave drive tones. For offsets of up to $\pm$\,200\,kHz, the Bell-state infidelity is still less than $10^{-2}$, demonstrating the robustness of the entangling operation to such offsets, or to differences between the frequencies of the qubits. The data in (B) were taken separately from the highest fidelity data reported in the main text, and the entangling operation fidelity was not fully optimized.}
\end{figure}

Our scheme requires a magnetic-field-sensitive qubit transition, so the qubit coherence time is limited by magnetic field fluctuations.  We investigate the performance of IDD, which should reduce the impacts of such fluctations, by observing the qubit coherence of a single ion in a spin-echo Ramsey experiment without applying the oscillating gradient. We compare two cases: either no fields are applied during the spin-echo arms, or we apply two microwave fields, symmetrically detuned about the qubit frequency by $\delta$, during both arms of the spin-echo.  The amplitudes of these fields are set to the IDD point $\Omega_\mu/\delta\approx0.601$, where $J_0\left(4\Omega_\mu/\delta\right)=0$ and the effects of low-frequency dephasing noise are thus suppressed~\cite{Sutherland2019}. Figure~\ref{fig_idd}A shows that including IDD during the spin-echo arms increases the spin-echo coherence time by more than an order of magnitude. Because the gradient at $\omega_g$ is not being applied, these microwave fields realize IDD but do not drive qubit-motion coupling.  

Our entangling operation ideally transforms the initial state $\ket{\downarrow\downarrow}$ to the symmetric $\ket{\Phi}\equiv\frac{1}{\sqrt{2}}(\ket{\downarrow\downarrow}+i\ket{\uparrow\uparrow})$ Bell state. We generate this entangling operation by applying the gradient and microwave fields as shown in Fig.~\ref{fig_general}A, using a sequence of eight pulses of simultaneously applied radiofrequency and microwave currents, interleaved with five qubit $\pi$ pulses, and a $\pi/2$ pulse at the beginning and end of the sequence (see Fig.~\ref{fig_pulse_sequence} in Ref.~\cite{supplementary}).  This entire operation has a total duration of 740\,$\mu$s~\cite{supplementary}. To achieve the highest fidelities, we sideband cool~\cite{Monroe1995} both the motional mode used for the entangling operation and the out-of-phase axial mode beforehand~\cite{supplementary}.  

\begin{figure}[h]
\includegraphics[width=0.9\textwidth]{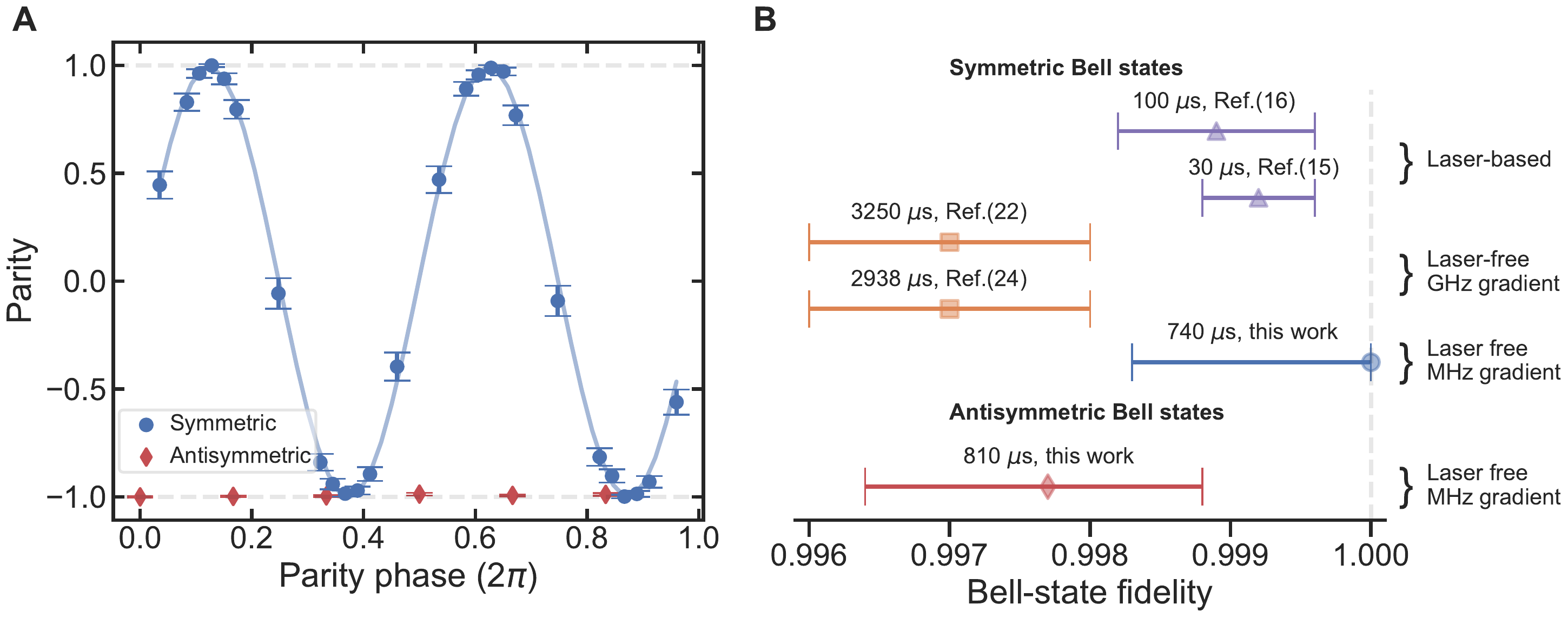}
\centering
\caption{\label{fig_parity}
 \textbf{(A)} Parity of symmetric and antisymmetric Bell states. After creating the Bell state, we apply a global $\pi/2$ analysis pulse with a variable phase, and then measure the resulting state's parity ${(P_0+P_2-P_1)}$.  The symmetric Bell state ${\ket{\Phi}=\frac{1}{\sqrt{2}}(\ket{\downarrow\downarrow}+i\ket{\uparrow\uparrow})}$ (blue circles) exhibits a sinusoidal parity oscillation (blue line shows best fit). We also create the antisymmetric state $\ket{\Psi_-}=\frac{1}{\sqrt{2}}(\ket{\downarrow\uparrow}-\ket{\uparrow\downarrow})$ using individual ion addressing. This state has a constant parity close to $-1$ (red diamonds), as it is invariant under global rotations. \textbf{(B)} Comparison of the highest-fidelity Bell states created in trapped ion qubits with laser-based and laser-free techniques. We plot corrected Bell state fidelities, as described in the respective references, and 68~\% confidence intervals, and list durations of the entangling operations. Results from the literature with lower Bell state fidelities than those presented here are not shown.}
\end{figure}

The fidelity of the prepared state is determined by a parity analysis method~\cite{Sackett2000, supplementary}. We measure the probabilities $P_0$, $P_1$, and $P_2$ of finding 0, 1, or 2 ions in $\ket{\downarrow}$, respectively, either immediately after the entangling sequence, or after a subsequent $\pi/2$ analysis pulse with a variable phase.  In the latter case, we determine the parity $P_0+P_2-P_1$ as a function of the analysis pulse phase as shown in Fig.~\ref{fig_parity}A.  We use these data to determine the Bell state fidelity using maximum likelihood estimation.

To characterize the performance of the entangling operation, we seek to estimate the fidelity with which the operation could create a Bell state from a pure unentangled input state.  As the experimental input states were not perfectly pure, we correct our reported Bell state fidelities for initialization outside the $\{\ket{\downarrow},\ket{\uparrow}\}$ manifold that occurs with probability $3.5(2)\times10^{-3}$ per qubit~\cite{supplementary}.  Due to statistical uncertainty in both the raw fidelity estimate (which is constrained to be between 0 and 1) and the estimate of the state initialization error, it is possible to calculate a corrected fidelity greater than 1, in which case we truncate the estimate to the physical maximum of 1.  We recorded multiple independent datasets while adjusting experimental parameters to optimize the fidelity. To avoid selection bias in choosing which dataset to report, we divided each dataset in half deterministically and used the extracted fidelity of one half as a ``trigger''.  The fidelities reported here are determined by selecting the dataset with the highest ``trigger'' fidelity and reporting the fidelity estimated only from the other half of that dataset. Analysis details are presented in the Supplementary Material~\cite{supplementary}.  

The estimated fidelity of the state produced (ideally $\ket{\Phi}$) for the dataset with the highest ``trigger'' fidelity, corrected for the initialization error, was 1.  From the distribution of bootstrapped fidelities, we determined a 68~\% confidence interval on the fidelity of [0.9983, 1] and a median bootstrap fidelity of 1. As an additional cross check, we calculated the fidelity using an unbiased linear estimator instead of the maximum likelihood parity analysis, obtaining consistent results~\cite{supplementary}. In Fig.~\ref{fig_parity}B, we compare the Bell-state fidelity and confidence interval to those of the highest-fidelity entangled states generated by laser-based~\cite{Gaebler2016, Ballance2016} and laser-free~\cite{Harty2016, Zarantonello2019} methods. We estimate that the leading sources of infidelity are decoherence of the ion motion such as motional dephasing, motional frequency drifts, and motional heating, giving a total estimated infidelity of approximately $7\times10^{-4}$, based on independent calibrations and numerical simulations~\cite{supplementary}.  These errors are consistent with the experimental results given the uncertainty in the fidelity estimate. The motional errors could in principle be reduced further by increasing the interaction strength, performing a gate sequence with more phase-space loops, or by using more complicated phase-space trajectories~\cite{Sutherland2020}. Future work will aim to reduce the uncertainty in the fidelity estimate and to characterize the entangling interaction using randomized benchmarking~\cite{Emerson2005, Knill2008}.

We also investigate the entangling operation's robustness to qubit frequency offsets that cause errors of the form $\frac{\hbar \epsilon}{2} \hat{\sigma}_{zi}$, which commute with the interaction in Eq.~\ref{eq_hamiltonian}. The $\pi$ pulses in the gate sequence and the IDD should both provide protection against such errors. To characterize this effect, we intentionally add a common offset to the frequencies of the detuned microwave fields with respect to the qubit frequency, then perform the entangling interaction and measure the Bell state fidelity, keeping all other parameters constant. As shown in Fig.~\ref{fig_idd}B, the Bell state infidelity remains below $10^{-2}$ for frequency offsets up to $\pm\,200$\,kHz.  

This insensitivity to qubit frequency offsets enables individual addressing of the ions in frequency space without compromising entanglement fidelity. Individual addressing of trapped ion qubits has been demonstrated previously using tightly focused laser beams~\cite{Naegerl1999, SchmidtKaler2003, Debnath2016, Wright2019, Erhard2019}, and without lasers using static magnetic field gradients~\cite{Johanning2009, Khromova2012, Piltz2014} and oscillating magnetic field gradients~\cite{Warring2013b,Craik2017} at GHz frequencies.  In our system, the currents at $\omega_g$ in the three qubit control electrodes give rise to a magnetic field with a strong spatial gradient, but nearly zero field amplitude along the quantization axis, at the ion positions.  The residual magnetic field at $\omega_g$ induces an ac Zeeman shift on the qubit transition frequency~\cite{supplementary}.  We apply static electric fields to rotate the ion crystal slightly with respect to the trap axis (Fig.~\ref{fig_general}B inset), such that the two ions experience different ac Zeeman shifts when the drive at $\omega_g$ is on; we choose the ion positions to give a roughly 20\,kHz differential shift.   This differential ac Zeeman shift produces an effective spin flip on a single qubit when using a spin-echo Ramsey sequence with a duration of approximately 70\,$\mu$s. By this individual control of our qubits, our initial symmetric entangled state $\ket{\Phi}$ is transformed into an antisymmetric entangled state $\ket{\Psi_-}$:

\begin{align}
    \ket{\Phi}=\frac{1}{\sqrt{2}}(\ket{\downarrow\downarrow}+i\ket{\uparrow\uparrow}) \Rightarrow \ket{\Psi_-}=\frac{1}{\sqrt{2}}(\ket{\downarrow\uparrow}-\ket{\uparrow\downarrow}).
\end{align}

\noindent After creating the antisymmetric state $\ket{\Psi_-}$, we measure the ion populations and parity as before.  Since $\ket{\Psi_-}$ is invariant under global rotations, we observe a constant parity as a function of the analysis pulse phase, as shown in Fig.~\ref{fig_parity}A. We use the same ``trigger'' technique as before to choose the reported dataset, which has a $\ket{\Psi_-}$ Bell state fidelity of 0.9977, again corrected for the initialization error on $\ket{\downarrow\downarrow}$; to the best of our knowledge, this is the highest reported fidelity in any platform for such an antisymmetric Bell state.  Bootstrapping yields a 68\% confidence interval for the Bell state fidelity of [0.9964, 0.9988], with a median bootstrap fidelity of 0.9976~\cite{supplementary}. We do not correct for any error in the individual addressing operation in this fidelity estimate. Imperfect calibration of the required duration and phase of the individual addressing operation, an effect that could be mitigated by use of a composite pulse sequence, may account for the reduced fidelity compared to our symmetric entangled state.

Our results highlight the potential of laser-free techniques for universal control of trapped ion qubits for quantum computing and simulation. This technology offers potential advantages for scaling of trapped ion quantum processors by enabling entangling operations to be carried out simultaneously in many trapping zones in a multi-zone ion trap, since the control currents can produce the necessary gradients and fields in multiple zones at the same time.  The entanglement of any particular group of ions could be enabled simply by adjusting the trap confinement in that zone with static potentials applied to local trapping electrodes, shifting the motional mode frequency of each zone in or out of resonance with the entangling interaction. The ability to perform many entangling operations simultaneously may reduce or eliminate the speed penalty relative to laser-based operations, which are often performed serially on different sets of ions due to laser power constraints.  The tolerance of our laser-free method to drifts or offsets in driving parameters, as well as other sources of decoherence, relaxes the requirements for the fields to be exactly the same across all zones, and the scheme does not rely on carefully tuned trap electrode dimensions to achieve high fidelity~\cite{Hahn2019, Zarantonello2019}. Changes in the local trapping potentials can also be used to select which ions are temporarily frequency-shifted for individual qubit control. Entanglement between qubits in different ion species could be achieved by adding another pair of weak microwave tones for each different qubit frequency; any magnetic-field-sensitive Zeeman or hyperfine qubit transition can be used.  These features may enable a new large-scale, multi-zone ion trap quantum computing architecture using a two-ion-species qubit/helper design, where all qubit operations aside from ion loading are carried out with radiofrequency or microwave signals, along with microwatt-scale laser beams for the helper species.  Laser-free mixed-species entanglement may also be useful for molecular~\cite{Chou2017} or highly charged ions\cite{Kozlov2018}, or trapped electrons~\cite{Matthiesen2021} or positrons, where suitable optical transitions for quantum logic operations may not be readily available.  

\section*{Acknowledgments}
We would like to thank C. J. Ballance, T. P. Harty, J. P. Gaebler, S. B. Libby, D. M. Lucas, V.~M.~Sch{\"a}fer, and T. R. Tan for helpful discussions. We thank M. Affolter and A.~L.~Collopy for helpful and insightful comments on the manuscript. At the time the work was performed, R.S., S.C.B., H.M.K., A.K., and D.T.C.A. were supported as Associates in the Professional Research Experience Program (PREP) operated jointly by NIST and the University of Colorado Boulder under Award No.\,70NANB18H006 from the U.S. Department of Commerce, National Institute of Standards and Technology. 
{\bf Funding:} This work was supported by the NIST Quantum Information Program and ONR. 
{\bf Author contributions:} R.S. and H.M.K. carried out the experiments, assisted by S.C.B., D.T.C.A, and D.H.S.; D.H.S., R.S., H.M.K., A.K., and R.T.S. analyzed the data and performed numerical simulations, with support from E.K. and S.G.; D.T.C.A., D.H.S., R.S., S.C.B., and H.M.K. built and maintained the experimental apparatus; R.S. wrote the manuscript with input from all authors; A.C.W., D.L., D.H.S., and D.J.W. secured funding for the work; and D.H.S. and D.T.C.A. supervised the work with support from A.C.W., D.L., S.G., E.K., and D.J.W.
{\bf Competing interests:} The authors declare no competing interests.

\noindent$^{\dagger}$Current address: Department  of  Physics,  University  of  Oxford,  Clarendon  Laboratory, Parks  Road,  Oxford  OX1  3PU,  U.K.\\
\noindent$^{\xi}$Current address: Department of Physics, Stanford University, Stanford, CA 94305.
\bibliographystyle{Science} 
\bibliography{lf_zz_refs}
\clearpage

\textbf{\Large Supplementary materials}

\setcounter{equation}{0}
\setcounter{figure}{0}
\renewcommand{\theequation}{S\arabic{equation}}
\renewcommand{\thefigure}{S\arabic{figure}}

\section{Details of experimental setup}

We operate in an externally applied magnetic field $|\vec{B}_0|=21.3$\,mT, where the qubit transition $\ket{\downarrow}\leftrightarrow\ket{\uparrow}$ has a magnetic field sensitivity of ${\frac{\partial\omega_0}{\partial B}/2\pi=-19.7}$\,MHz/mT.  The value of $|\vec{B}_0|$ was chosen to provide a first-order magnetic-field-insensitive qubit transition between the states $\ket{\downarrow^\prime}\equiv{^2S}_{1/2}\ket{F=3,m_F=1}$ and $\ket{\uparrow^\prime}\equiv{^2S}_{1/2}\ket{F=2,m_F=1}$, with a frequency of 1.686~GHz. Such field-insensitive qubits can have measured coherence times of many seconds\cite{Harty2014} or even minutes~\cite{Wang2021}. When not performing entangling gates, we could map the populations in $\ket{\downarrow}$ and $\ket{\uparrow}$ into the states $\ket{\downarrow^\prime}$ and $\ket{\uparrow^\prime}$ using microwave control pulses, allowing the qubit state to be stored coherently for much longer times than is possible in the $\{\ket{\downarrow},\ket{\uparrow}\}$ qubit.

We prepare the $\ket{\downarrow}$ state by optical pumping on the ${^2S_{1/2}\leftrightarrow {^2}P_{3/2}}$ transitions at 280\,nm with $\sigma^+$ polarized light. During and after sideband cooling, further optical pumping pulses on these transitions are applied to repump population into $\ket{\downarrow}$.  Qubit readout is performed by detecting fluorescence from the laser-driven $\ket{\downarrow}\leftrightarrow\ket{^2P_{3/2}, F=4, m_F=4}$ cycling transition. To reduce readout errors, we use microwave pulses to ``shelve" the $\ket{\uparrow}$ state to the $\ket{^2S_{1/2}, F=2, m_F=-1}$ state, detuned by $\approx1.6\,$GHz from $\ket{\downarrow}$, before applying the readout laser beam.  Our readout can only distinguish the total number of ions fluorescing (0, 1, or 2), not the state of each individual ion.  We produce an oscillating magnetic field gradient at $\omega_g=2\pi\times5$\,MHz with an amplitude of 152(15)\,T/m at the ion by applying currents of 0.8(1), 1.1(1), and 1.1(1)\,A on qubit control electrodes 1, 2, and 3 respectively, shown in Fig.~\ref{fig_general}. The relative amplitudes and phases of these currents are chosen to minimize residual magnetic fields at $\omega_g$ at the ion position~\cite{Srinivas2019}; in practice, this means that the current in electrode 2 is driven approximately 180 degrees out of phase with the currents in electrodes 1 and 3.  These currents dissipate $\approx100\,$mW total in the trap from resistive losses.  The weaker currents used to generate the microwave-frequency magnetic fields dissipate $\approx$\,6\,mW in the trap. Currents in qubit control electrode 1 on resonance with the qubit frequency are used for global single-qubit operations. Based on sequences of 200 repeated $\pi$ pulses on the $\ket{\downarrow}\leftrightarrow\ket{\uparrow}$ transition using a single ion, the error per $\pi$ pulse has an upper bound of approximately $10^{-4}$. We generate the radiofrequency and microwave currents with high-speed digital-to-analog converters followed by commercially available amplifiers and filters, as described in Section~\ref{sec:electronics}.

To achieve the highest entangled state fidelities, the out-of-phase radial mode used for entangling operations is first cooled near its ground state with sideband transitions driven by the radiofrequency gradient and microwave tones, interleaved with optical repumping~\cite{Monroe1995, Srinivas2020}. This motional mode is coupled weakly to the out-of-phase axial mode; thermal phonon occupation in the out-of-phase axial mode will cause dephasing of the motional mode used for entanglement~\cite{Roos2008nonlinear, Nie2009}.  To reduce this dephasing, we also cool the out-of-phase axial mode near its ground state. For this mode, we use laser-based Raman sideband transitions, as the trap geometry prevents direct laser-free sideband cooling of the axial modes.

\section{Fields and gradients for entangling interaction}

The entangling interaction relies on the gradient of a magnetic field $\vec{B}_g$ oscillating at $\omega_g$, as well as a microwave magnetic field $\vec{B}_\mu$ with frequency components at $\tilde{\omega}_0\pm\delta$ (see Fig.~\ref{fig_general}A).  This latter field can be equivalently described as a microwave field at $\tilde{\omega}_0$ whose amplitude varies sinusoidally with time $t$ as $\cos(\delta t)$.  A full derivation of how the entangling interaction arises from these fields is given in Ref.~\cite{Sutherland2019}. As shown in Eq.~\ref{eq_hamiltonian}, the slowly-rotating terms generate the interaction

\begin{align}
\hat{H}_I(t) = \hbar\Omega_g J_2\left(\frac{4\Omega_\mu}{\delta}\right)(
\hat{\sigma}_{z1}-\hat{\sigma}_{z2})(\hat{a}e^{i\Delta t}+\hat{a}^\dagger e^{-i\Delta t}).  
\end{align}

\noindent The quantity $\Omega_g$ characterizes the strength of the gradient at $\omega_g$ used in the entangling interaction and is given by

\begin{equation}
\label{eq:Omegag}
\Omega_g \equiv \frac{r_0 [\nabla (\vec{B}_g\cdot\hat{r}_q)\cdot\hat{r}]}{4}\;\frac{\partial\omega_0}{\partial B} \Big|_{B=|\vec{B}_0|},
\end{equation}

\noindent where $\hat{r}$ is a unit vector along the motional mode, $\hat{r}_q$ is a unit vector along the quantization axis defined by an external static bias magnetic field $\vec{B}_0$, and $\frac{\partial\omega_0}{\partial B}$ is the magnetic field sensitivity of the qubit transition frequency. For these experiments $|\vec{B}_0|\approx21.3$\,mT, and $\vec{B}_0$ is in the plane of the trap electrodes at an angle of $67.5^\circ$ to the trap axis. The ground state extent of the ions' motional mode in the trap is $r_0=\sqrt{\hbar/2M\omega_r}$, where $\omega_r$ is the motional mode frequency and $M$ is the total mass of both ions. We use the expression in Eq.~\ref{eq:Omegag} to determine the strength of the magnetic field gradient from the entangling operation duration.  

The microwave fields are characterized by the corresponding resonant Rabi frequency 

\begin{equation}
\Omega_\mu\equiv\frac{B_x}{2\hbar}\bra{\downarrow}\hat{\mu}_x\ket{\uparrow},
\end{equation}

\noindent where $B_x$ is the component of the oscillating microwave magnetic field $\vec{B}_\mu$ perpendicular to the quantization axis and $\hat{\mu}_x$ is the magnetic moment of the ion in the same direction.

We can decompose $\vec{B}_g$ into $\vec{B}_{g,\parallel}+\vec{B}_{g,\perp}$, where $\vec{B}_{g,\parallel}\equiv(\vec{B}_g\cdot\hat{r}_q)\hat{r}_q$ is the component of $\vec{B}_g$ along the quantization axis and $\vec{B}_{g,\perp}$ is the component perpendicular to the quantization axis.
To maximize the entangling interaction strength and reduce the modulation of the qubit frequency~\cite{Srinivas2019}, we null $\vec{B}_{g,\parallel}$ at the ion positions by adjusting the relative phases and amplitudes of the currents at $\omega_g$ in all three qubit control electrodes.  However, $\vec{B}_{g,\perp}$ is not simultaneously nulled and gives rise to an ac Zeeman shift $\Delta_{ac}$ on the qubit frequency ~\cite{Srinivas2019}.  We have ${\Delta_{ac}/2\pi\approx-400}$\,kHz for the configuration of currents at $\omega_g$ used in the entangling operation. The qubit frequency in the absence of any control fields at $\omega_g$ is $\omega_0$. With the control fields at $\omega_g$, the qubit frequency is shifted to $\tilde{\omega}_0$, where $\tilde{\omega}_0=\omega_0+\Delta_{ac}$.

Like $\vec{B}_{g,\parallel}$, $\vec{B}_{g,\perp}$ has a spatial gradient, which means $\Delta_{ac}$ likewise has a spatial gradient.  If the ion crystal is rotated relative to the trap axis, the two ions will experience different values of $\vec{B}_{g,\perp}$ and thus different ac Zeeman shifts, which we describe in terms of the average shift $\Delta_{ac}$ on both ions and the differential shift $\delta_{ac}$ between the two ions. We use this differential shift $\delta_{ac}$ for individual addressing in frequency space.  In practice, we can achieve $\delta_{ac}/2\pi\approx20$\,kHz, with $\Delta_{ac}/2\pi\approx2.5\,$MHz, by driving a current at $\omega_g$ through only electrode 1.  In this single-electrode driving configuration, $\vec{B}_{g,\parallel}$ is not nulled, but this is not necessary during the individual addressing operation.  Choosing not to null $\vec{B}_{g,\parallel}$ allows higher values of $\delta_{ac}$ to be achieved.

\subsection{Radiofrequency and microwave drive electronics\label{sec:electronics}}

\begin{figure}
    \centering
    \includegraphics{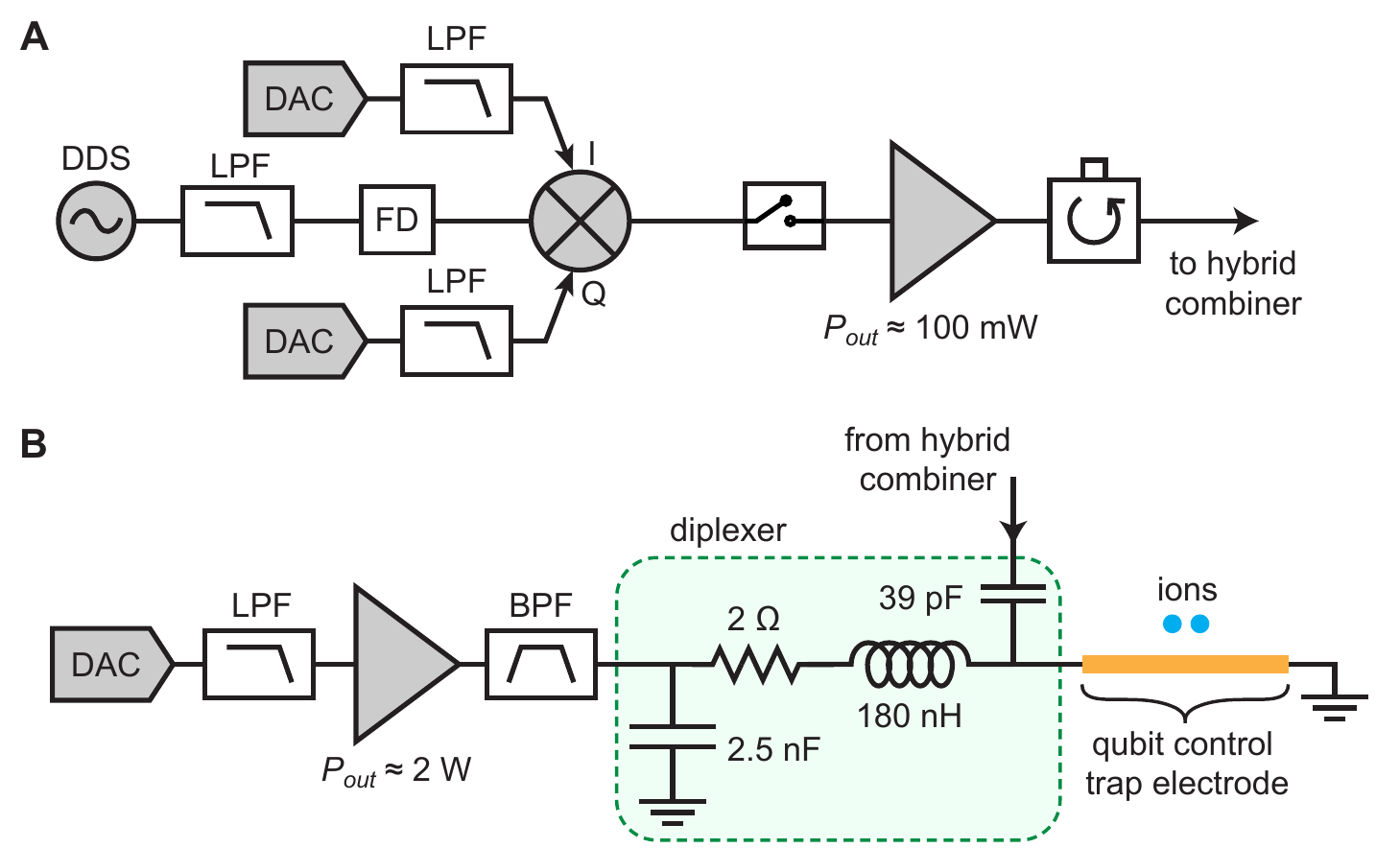}
    \caption{Drive electronics for qubit control fields.  Panel (\textbf{A}) shows a schematic of the microwave signal generation chain for a single frequency.  Two such chains are used, one operating at $\tilde{\omega}_0+\delta$ and the other at $\tilde{\omega}_0-\delta$, and are combined using a hybrid that combines signals at different frequencies. Panel (\textbf{B}) shows a schematic of the circuit used for generating radiofrequency control signals at $\omega_g$, combining them with the microwave signals, and delivering both to the trap.  The circuit shown is for a single trap electrode; two duplicate circuits drive the other two qubit control trap electrodes, each with a different phase and amplitude of the drive at $\omega_g$.  However, microwave signals are only sent to the trap on one of the three electrodes.  Abbreviations: LPF low-pass filter, BPF band-pass filter, FD frequency doubler, $P_\mathit{out}$ amplifier output power.}
    \label{fig:electronics}
\end{figure}

A schematic of the radiofrequency and microwave electronics used to generate the control fields is shown in Fig.~\ref{fig:electronics}.  The signals at $\omega_g/2\pi=5$\,MHz in the qubit control electrodes are generated by three independent 16-bit digital-to-analog converters (DACs), one per electrode, operating at 100\,MS/s.  The waveform frequencies, phases, and time-dependent amplitude envelopes are defined digitally, and then the waveforms are directly synthesized without subsequent analog modulation components.  The DAC chips are on a common circuit board and have a shared clock to ensure a deterministic phase relationship between the signals in all three qubit control electrodes~\cite{Bowler2013}.  The DAC outputs are each low-pass-filtered, amplified to roughly 2\,W per electrode, band-pass-filtered to remove harmonics and low-frequency noise, and sent to the low-frequency port of a custom resonant diplexer circuit (one per qubit control electrode) shown in Fig.~\ref{fig:electronics}B.  The diplexers provide narrowband impedance transformation for the tones at $\omega_g$, increasing the current in the trap electrodes for a given drive power by roughly a factor of 3 while reducing back-reflection to the amplifiers and providing additional bandpass filtering.  The diplexers also serve to combine the tones at $\omega_g$ with the microwave tones near $\omega_0$ onto the same trap electrodes.  In practice, while tones at $\omega_g$ are used on all three electrodes, the microwave tones are only applied to one electrode (see Section~\ref{sec:pulseseq} for details).  Most of the drive power at $\omega_g$ is dissipated in the resistive parts of the diplexer circuits; only about 100\,mW of the $\approx6 W$ of total drive power among all three electrodes is dissipated in the trap, due to resistive losses.  Improvements to the resonant diplexer circuit design could provide increased currents at $\omega_g$ in the trap, and thus faster entangling interactions, with the same or smaller drive power.  The qubit control electrodes are shorted to ground at the far end of the trap, approximately 3\,mm away from where the ions are confined, producing a standing wave with a current antinode near the ion location.  This doubles the current in the qubit control electrodes for a given drive power relative to using a $50\,\Omega$ termination and shifts most dissipation of the control signals to the impedance matching circuits and circulator terminations outside the vacuum system.

The microwave signals near $\omega_0/2\pi\approx1.326$\, GHz are generated by a pair of high-speed direct digital synthesizer (DDS) chips clocked at 2.4\,GS/s.  The DDS chips have a shared clock and are phase-synchronized to each other as well as to the DACs that generate the signals at $\omega_g$.  We filter, frequency-double, and amplify the DDS outputs, then use an IQ modulator on each channel to shape the amplitude envelope in time.  The I and Q ports of the modulator are driven by low-pass-filtered DACs of the same design as those used to generate the signals at $\omega_g$.  The microwave signals then pass through a microwave switch to provide increased on/off ratio compared to the modulator alone, after which they are amplified to $\sim100$\,mW.  The two tones are combined using a microwave hybrid and sent to one of the trap electrodes via the diplexer circuit described above, which combines them with the strong tone at $\omega_g$ for that electrode. A similar microwave chain, but without the IQ modulator, is used to generate rectangular microwave pulses at $\omega_0$ for performing global single-qubit rotations, which are coupled onto the qubit control electrodes using a directional coupler between the hybrid coupler and the diplexer.

\subsection{Entanglement pulse sequence\label{sec:pulseseq}}

\begin{figure}
    \centering
    \includegraphics[width=1\textwidth]{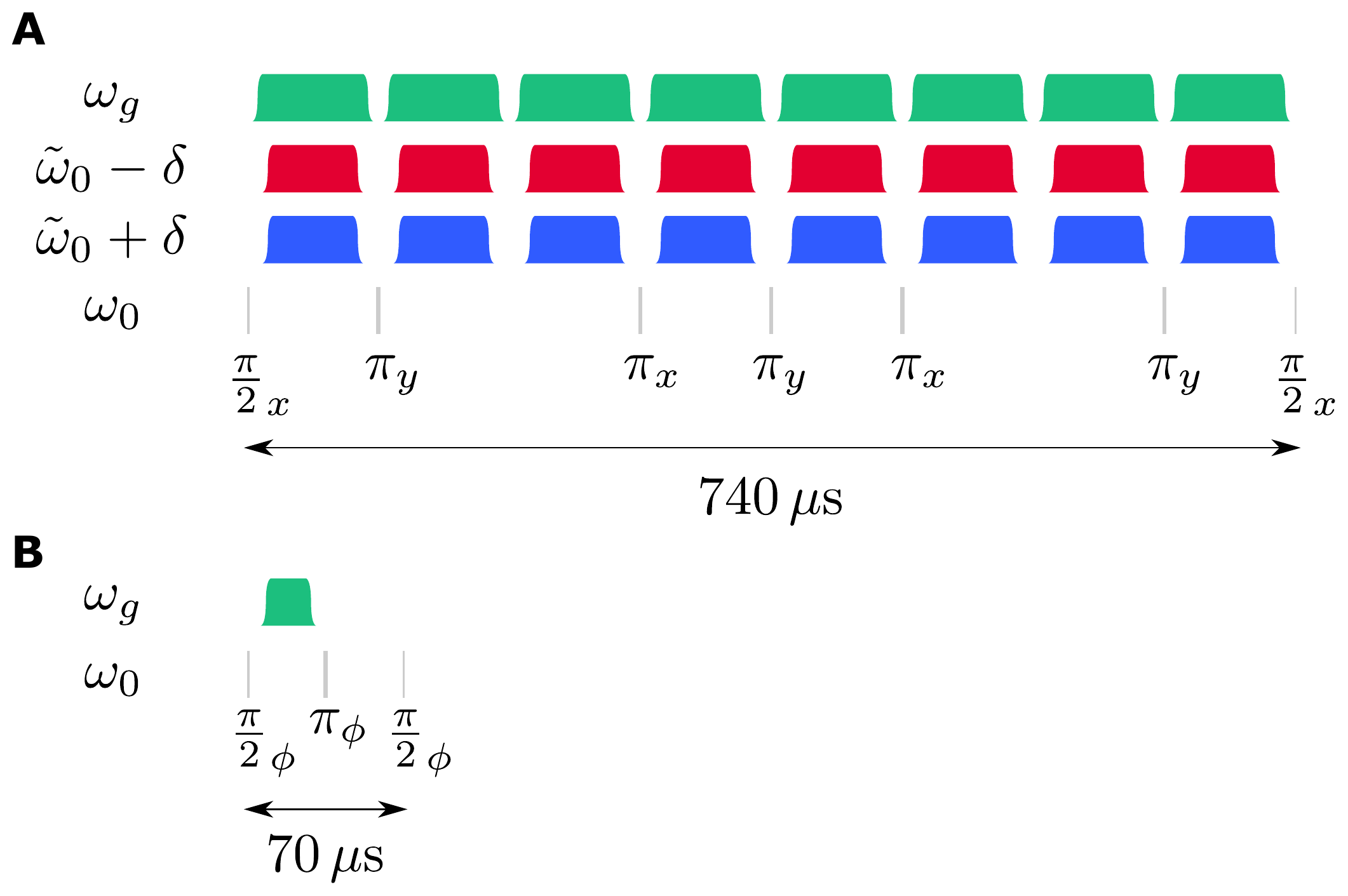}
    \caption{\textbf{(A)} Pulse sequence for entangling interaction. We plot the amplitude of the control signals schematically versus time.  The gradient oscillating at $\omega_g$ (green), as well as the two microwave fields symmetrically detuned by $\delta$ from the ac Zeeman shifted qubit frequency $\tilde{\omega}_0$ (red and blue), are ramped up and down eight times. Additional $\pi/2$ and $\pi$ pulses (gray) correct for qubit frequency offsets and drifts and perform the Walsh modulation. These pulses are at $\omega_0$, the qubit frequency unshifted by the ac Zeeman shift from the oscillating magnetic field at $\omega_g$. The total duration of our entangling interaction is 740\,$\mu$s. (\textbf{B}) Pulse sequence for single ion addressing. We perform a spin-echo Ramsey sequence, using the gradient oscillating at $\omega_g$ in only the first arm. The phase of the global qubit rotation pulses in this sequence, here denoted by $\phi$, must be calibrated relative to the entangling interaction in $\textbf{(A)}$ to create the desired state $\ket{\Psi_-}$.}
    \label{fig_pulse_sequence}
\end{figure}

We show the pulse sequences used for the gradient and microwave fields used to generate our symmetric and antisymmetric entangled states in Fig.~\ref{fig_pulse_sequence}.  The currents at $\omega_g$ and $\tilde{\omega}_0\pm\delta$ are ramped up and down smoothly over 5 $\mu$s, with rising and falling edges following an approximate sine-squared envelope.  Due to the size of the ac Zeeman shift $\Delta_{ac}$, the drive at $\omega_g$ is ramped up completely before the microwave currents at $\tilde{\omega}_0\pm\delta$ are ramped up, and the downward ramps are carried out in reverse order.  The $\pi/2$ and $\pi$ pulses at $\omega_0$ are carried out when all the other currents are turned off, so the qubit frequency is not ac-Zeeman-shifted and the qubit is not driven off-resonantly.  The phases of these pulses, denoted with subscripts $x$ and $y$, where $y$ indicates a $90^\circ$ phase shift with respect to $x$, are chosen to provide robustness to miscalibrations (overrotation or underrotation) in the $\pi$ pulses.

The total duration of the entangling operation is 740\,$\mu$s, of which the up and down ramps (during which relatively little entanglement is generated) consume 160\,$\mu$s.  Increased currents at $\omega_g$ would increase $\Omega_g$ and thus reduce the overall gate duration.  Decreasing the durations of the ramps must be done with care, as shorter ramp durations can lead to increased off-resonant qubit transitions that reduce the fidelity of the entangling operation~\cite{Sutherland2019}.  

\section{Fidelity analysis}

In this section, we describe the methods used for Bell state fidelity analysis, including the correction of state preparation and measurement (SPAM) errors.  

\subsection{Estimating SPAM errors}

We use reference data to determine our state preparation and measurement (SPAM) errors. We repeatedly prepare the ions approximately in the $\ket{\downarrow\downarrow}$ state and measure the ion fluorescence, building a histogram of the number of photon counts observed during the detection period.  We also perform the same experiment but with a microwave $\pi$ pulse at $\omega_0$ applied to the ions prior to measurement, taking multiple sets of about 18,500 measurements per state.  These experiments give reference count histograms for two or zero ions fluorescing, respectively (two ``bright'' ions or two ``dark'' ions).  These are approximately Poissonian-distributed but contain additional corrections due to off-resonant pumping of dark ions into the bright state (``repumping''), optical pumping of bright ions into the dark state due to imperfect closure of the cycling transition (``depumping''), and imperfect optical pumping in the state preparation process that leaves an ion in a state outside the $\{\ket{\downarrow},\ket{\uparrow}\}$ manifold (``leakage'')~\cite{Srinivas2020}.  We use maximum likelihood (ML) to estimate the Poissonian mean count rates from bright and dark ions, as well as depumping, repumping and leakage rates.  We use these rates to construct calculated reference distributions for the case of zero, one, and two ions fluorescing.  

We determine $P_0$, $P_1$, and $P_2$ after an entangling operation using ML estimation based on the constructed reference distributions.  According to tests using simulated count histogram data, the resulting state measurement error is at the $\approx10^{-4}$ level or better for reference datasets of the size used in our fidelity analysis.  

Using this technique, we determine the rate of state initialization errors due to leakage (note that this leakage occurs during preparation of the initial $\ket{\downarrow\downarrow}$ state, not during the entangling operation itself).  Our reported fidelities are all corrected for this imperfection in state initialization, as described below.  For the symmetric Bell state data, we determined a state initialization error rate per qubit due to leakage of $3.5(2)\times10^{-3}$, while for the antisymmetric Bell state data, taken several weeks later, this rate was $1.7(2)\times10^{-3}$.  It is not possible using our analysis to distinguish between an ion incorrectly initialized in the $\ket{\uparrow}$ state and an error in the microwave $\pi$ pulse; however, the data bound the combined rate of these two errors to be at or below $\approx 10^{-4}$ per qubit.  We do not attempt to correct for either of these two types of errors, both of which would decrease the fidelity of the final entangled state.  

\subsection{Parity and population calculation\label{sec:parpop}}

Each dataset consists of 40 sets each containing 200 repetitions of the entangling operation followed by population measurement (``population'' trials) and 52 sets each containing 200 repetitions of the entangling operation followed by parity analysis pulses, with a different analysis phase for each set (``parity'' trials).  For the antisymmetric Bell states, we use 42 sets of 200 repetitions each as parity trials, composed of 7 sets at each of 6 evenly spaced phase values.  We divide each dataset into ``trigger'' and ``analysis'' halves by assigning half of the population trials from each set (and for the antisymmetric Bell state, half of the parity trials from each set), chosen at random, to the ``trigger'' subset.  For the symmetric Bell state, we assign every other set of 200 parity trials (versus analysis phase) to the ``trigger'' subset, rather than splitting each set of 200 in half.  This choice improved the parity estimation based on analysis of simulated data.  The random assignment algorithm is seeded by a hash of the unique dataset identifying number.  

The fidelity of the final Bell state is determined from the populations and parity as described below.  The populations are determined by combining all population trials into a single histogram and estimating $P_0$, $P_1$, and $P_2$ using maximum likelihood based on the reference distributions described above.  For the antisymmetric Bell state, the parity is also determined by combining all parity trials into a single histogram and estimating $P_0$, $P_1$, and $P_2$ in the same way.  The values of the parity analysis phases in this instance were chosen as multiples of $\pi/3$, such that the extracted parity using this method would average over the amplitude of any parity oscillations due to residual population in a symmetric Bell state.  

To determine the amplitude of parity oscillations for the symmetric Bell state, we maximize the joint likelihood over the count histograms for all the different analysis phases, assuming a parameterized model of a sinusoidal parity oscillation versus phase and employing the reference count distributions described above.  

\subsection{Effects of initialization error}

The density matrix describing the state of each ion $\rho_{0, 1}$ after initialization is 

\begin{align}
    \rho_{0, 1} = (1-\epsilon)\ket{\downarrow}\bra{\downarrow}+ \epsilon\ket{a}\bra{a},
\end{align}

\noindent where $\epsilon$ is the probability of producing the leaked state $\ket{a}$. The state $\ket{a}$ is outside of the qubit manifold and does not participate in the gate dynamics. This initialization error most likely occurs due to photon scattering from our Raman beams into the $\ket{F=3, m_F=1}$ state. This state does not fluoresce and is detected as dark through our detection process. We ignore any initialization in the $\ket{\uparrow}$ state, since our reference data indicate that this process is at least an order of magnitude less likely than producing the state $\ket{a}$. Furthermore, our reference data cannot distinguish preparation in the $\ket{\uparrow}$ state from an imperfect $\pi$ pulse on the $\ket{\downarrow}\rightarrow\ket{\uparrow}$ transition. As any population in $\ket{\uparrow}$ state would only reduce the fidelity of the entangled state we measure, we do not correct for this error. For two ions, the initial density matrix is

\begin{align}
\label{eq_8_start_state}
    \rho_{0, 2} = (1-\epsilon)^2\ket{\downarrow\downarrow}\bra{\downarrow\downarrow} + \epsilon(1-\epsilon)\big(\ket{\downarrow a}\bra{\downarrow a} + \ket{a \downarrow}\bra{a \downarrow}\big) + \epsilon^2\ket{aa}\bra{aa}.
\end{align}

\noindent We now analyze the effect of this initialization error on both the symmetric and antisymmetric entangled state fidelities. 

\subsubsection{Symmetric entangled state fidelity}

Our entangling operation acts on a perfect $\ket{\downarrow\downarrow}$ input state to produce a symmetric entangled state ${\ket{\Phi}\equiv\frac{1}{\sqrt{2}}(\ket{\downarrow\downarrow}+i\ket{\uparrow\uparrow})}$ with a fidelity $F$. The same operation acting on $\ket{\downarrow a}$ and $\ket{a \downarrow}$ instead flips the spin of the ion within the qubit manifold, resulting in $\ket{\uparrow a}$ and $\ket{a \uparrow}$ respectively, to first order. There will be higher order terms that scale as $(1-F)\epsilon$, which we ignore. The entangling operation has no effect on the $\ket{a a}$ state. To determine the entangling operation fidelity $F$ from the measured fidelity $F_m$, we must correct for the initialization error caused by leakage. We determine the fidelity $F_m$ of our entangled state by measuring the ion populations after our entangling operation and the parity with an additional $\pi/2$ pulse with a variable phase\cite{Sackett2000}. The leaked states have no effect on the parity measurement, but will increase the measured populations as the leaked states are indistinguishable from $\ket{\uparrow\uparrow}$. Thus, $F_m$ is

\begin{align}
\begin{split}
    F_m &= (1-\epsilon)^2F+\epsilon(1-\epsilon)+\frac{\epsilon^2}{2} \\
    &= F-\epsilon(2F-1)+\epsilon^2(F-\frac{1}{2}) \\
    &\approx F-\epsilon,
\end{split}
\end{align}

\noindent for $(1-F), \epsilon \ll 1$.

\subsubsection{Antisymmetric entangled state fidelity}

As with the symmetric state, we estimate the initialization-corrected fidelity $F$ by making measurements of the populations after (ideally) generating the antisymmetric state $\ket{\Psi_-}=\frac{1}{\sqrt{2}}(\ket{\downarrow\uparrow}-\ket{\uparrow\downarrow})$ and the parity with an additional $\pi/2$ pulse. Unlike the symmetric state, the ideal antisymmetric state has parity -1 for all phases. The parity for all other states is greater than -1. Thus, instead of trying to fit a low amplitude oscillation, we instead take the average parity from measurements with six different phases of the $\pi/2$ pulse equally spaced from 0 to 2$\pi$\cite{Srinivas2020}. The symmetric state has 0 parity averaged over these phases..

For the leaked state, the single-ion addressing sequence, to first order, will transform ${\ket{\uparrow a}\rightarrow\ket{\downarrow a}}$ while leaving the $\ket{a\uparrow}$ state unchanged. The parity of these states after applying an additional $\pi/2$ pulse is 0,  but these states will contribute to the population measurements. For the antisymmetric state, the measured fidelity $F_m$ is related to F by

\begin{align}
\begin{split}
    F_m &= (1-\epsilon)^2F+\frac{\epsilon}{2}(1-\epsilon) \\
    &= F-\epsilon(2F-\frac{1}{2})+\epsilon^2(F-\frac{1}{2}) \\
    &\approx F-\frac{3}{2}\epsilon,
\end{split}
\end{align}

\noindent again for $(1-F),\epsilon\ll1$.

\subsection{Bootstrap analysis}

\begin{figure}
    \centering
    \includegraphics[width=0.9\textwidth]{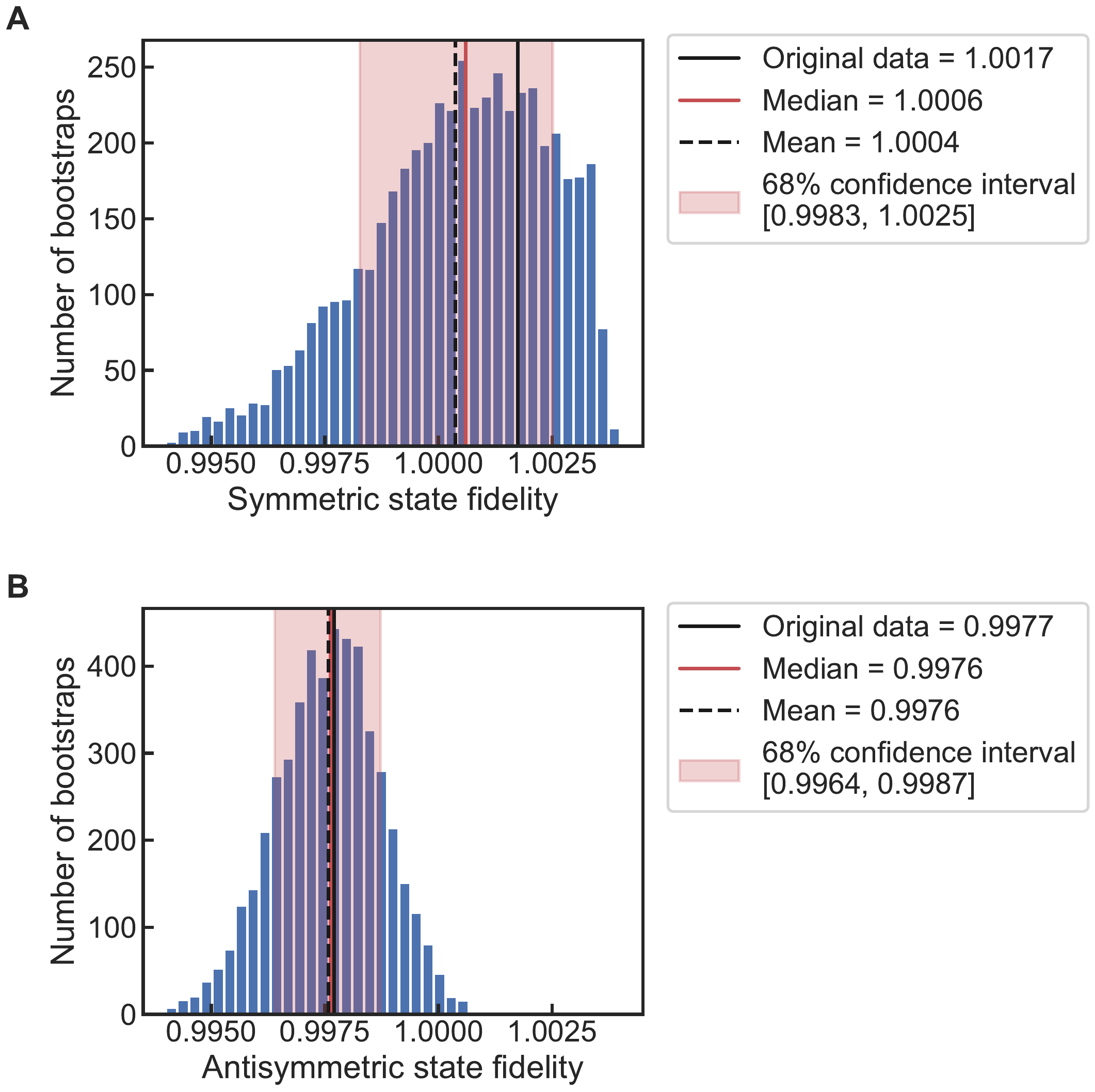}
    \caption{We perform a bootstrap analysis on our symmetric (\textbf{A}) and antisymmetric entangled state data (\textbf{B}) to determine the confidence intervals of Bell-state fidelities, corrected for state initialization errors. For each state, we resample the counts of the data with replacement 5000 times and reanalyze the data. For the means and 68\% confidence intervals reported in the text, we truncate all fidelities at 1. \textbf{(A)} Bootstrapped fidelities for the symmetric state.  (\textbf{B}) Bootstrapped fidelities for the antisymmetric state. }
    \label{fig_bootstraps}
\end{figure}

To generate the confidence intervals on our entangled state fidelities, we generate 5,000 synthetic (``bootstrapped") data sets by resampling the experimental photon count data for the entangled state analysis with replacement~\cite{Efron1994}. We also resample the corresponding reference data 5,000 times, each bootstrap having corresponding estimates of Poissonian means, depumping, repumping, and leakage.  We construct the corresponding reference count distributions using additional variation of the Poissonian means, based on the magnitude of slow variations in these means seen experimentally over the course of a day.  We then analyze each bootstrapped data set using a bootstrapped reference count distribution.  The distributions of the resulting fidelities, corrected for preparation errors, for the triggered symmetric and antisymmetric Bell states are shown in Fig.~\ref{fig_bootstraps}A and B, respectively.

Due to statistical uncertainty in both the estimate of the uncorrected fidelity and the initialization error, our fidelity analysis method can give corrected fidelity estimates that exceed 1, which are nonphysical.  When reporting fidelities, we therefore truncate any estimated fidelities greater than 1 at a value of 1. The lower and upper endpoints of our confidence intervals are the 16th and 84th percentile points of the distribution of bootstrapped fidelities, truncated to 1 as necessary. Note that the confidence interval for the symmetric state fidelity has both median and 84th percentile values of 1.

\subsection{Dataset selection}

The data were taken before the detailed fidelity analysis tools presented in this manuscript were developed, so the experimental parameters were adjusted using less accurate fidelity estimation techniques to guide the optimization.  Because of the inability to characterize and compensate for very small infidelities on the fly, the data were taken while sweeping experimental parameters (pulse durations and amplitudes/frequencies of control signals) through a range of values near the predicted optimum.  As a result, some of the datasets were taken with nonoptimal parameters and should have lower-than-optimal underlying Bell state fidelity, but there is not an a priori method of knowing which datasets are which.  Furthermore, drifts in both the motional frequency and the qubit frequency can cause nonmonotonic variation in the underlying Bell state fidelity due to effects related to the specific parameters of the shaped pulse rise and fall profiles~\cite{Srinivas2020}.  There may also be other uncharacterized system parameters whose drifts cause fluctuations in the underlying Bell state fidelity, despite performing the entangling interaction with nominally identical parameters.  In addition to any variation in underlying Bell state fidelity, there is statistical uncertainty (and thus dataset-to-dataset variation) in estimating the underlying Bell state fidelity. Each set of entanglement data took roughly 7 minutes to acquire.

\begin{figure}
    \centering
    \includegraphics[width=0.5
    \textwidth]{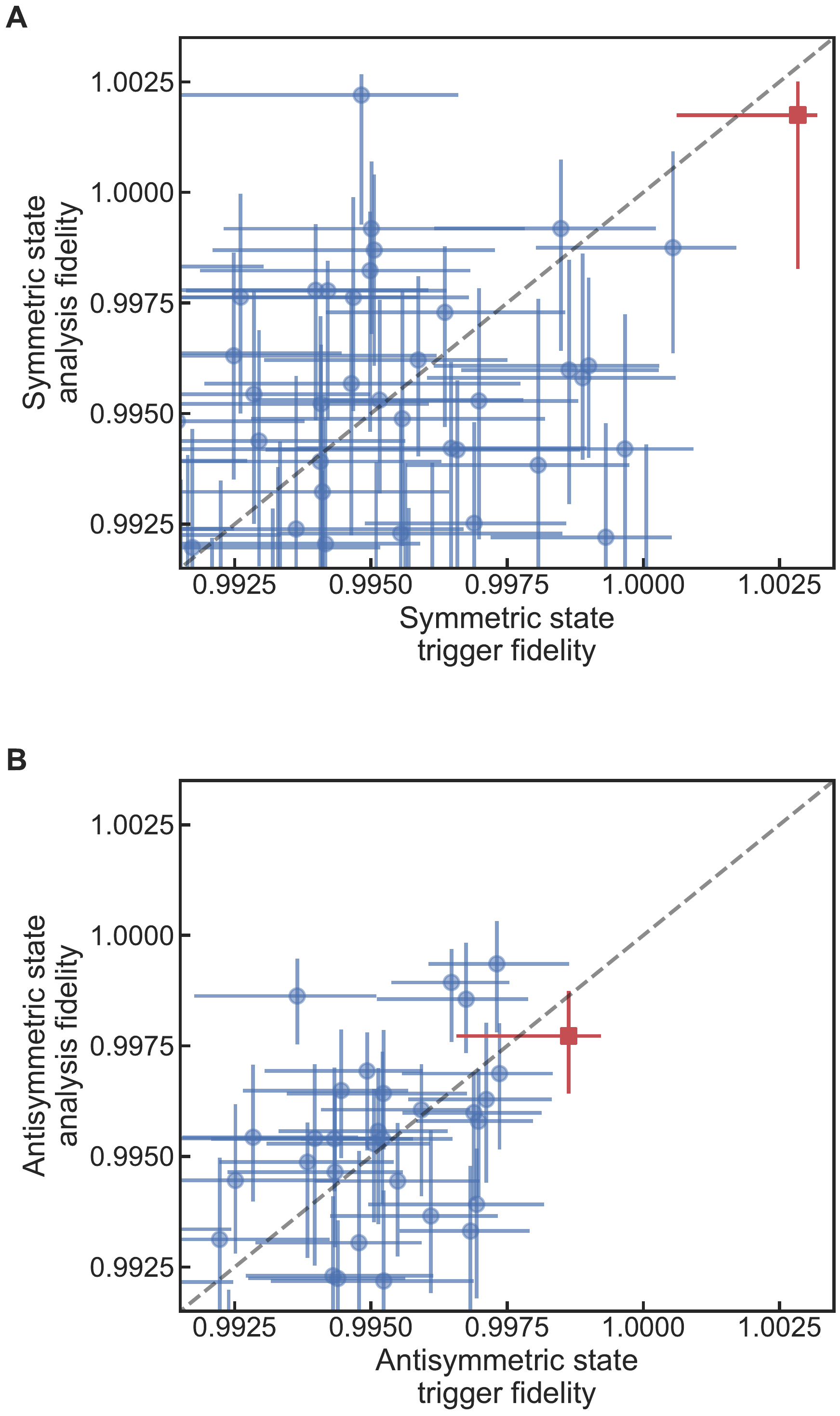}
    \caption{We use a ``trigger" to select both the symmetric (\textbf{A}) and antisymmetric entangled state data (\textbf{B}) to report. For each set of data, we split each of our datasets into two. We use half to compute a ``trigger" fidelity, and the other half to analyze the fidelity of the created state. We use the dataset with the highest trigger fidelity. The data used to determine the trigger fidelity are subsequently discarded and are not used to compute the gate fidelity. This process avoids relevant selection bias in choosing which fidelity to report. The red square denotes the dataset with the highest trigger value which we report in the main text. The error bars are determined by nonparametric bootstrapping, indicating the 68\% confidence interval. Dashed lines indicate a slope of 1.}
    \label{fig_triggers}
\end{figure}

As a result, we cannot a priori determine a single dataset or set of datasets that are anticipated to represent the best underlying Bell state fidelity.  In order to avoid selection bias in determining which datasets to report as our best dataset, we use independent ``trigger'' data.  We divide every set of experimental data, which includes both the parity and population measurements, in half, as described in Section~\ref{sec:parpop}. We use one half of the data as the ``trigger," and report the fidelity of the other ``analysis'' half of the dataset with the highest ``trigger" fidelity. Crucially, the ``trigger'' data are only used to select which dataset's ``analysis'' half to report; they are not used when calculating the fidelity of the ``analysis'' half. The ``trigger'' and ``analysis'' fidelities from the datasets, for both the symmetric and antisymmetric entangled states, are shown in Fig.~\ref{fig_triggers}.  

\subsection{Linear estimator}
We also use an unbiased linear estimator of Bell-state fidelity as a consistency check, implementing techniques similar to the linear estimation of process fidelity in \cite{Wan_2019} and the linear state tomography described in \cite{Zhu_2014} that have been adapted for this analysis.

For an abstract experiment consisting of $j$ different positive operator-valued measure (POVM) measurements $\{\Pi_i^{(j)}\}$ of an unknown state $\rho$, a linear estimator of the fidelity of $\rho$ with a target state $\ket{\psi}\bra{\psi}$ can be constructed from any decomposition of $\ket{\psi}\bra{\psi}$ into a linear combination of the POVM elements $\{\Pi_i^{(j)}\}$. Such a decomposition is described by a set of real coefficients $\alpha_i^{(j)}$ that satisfy

\begin{equation}\label{le_coefficient_condition}
\ket{\psi}\bra{\psi} = \sum_i\sum_j \alpha_i^{(j)} \Pi_i^{(j)}.
\end{equation}

\noindent In this specific experiment, the different POVM measurements indexed by $j$ correspond to the different collective rotations applied to the state before measurement. For each fixed $j$, the $i$th POVM element represents the measurement outcome associated to the observation of $i$ photons during the measurement. The POVMs satisfy the standard normalization $\sum_i \Pi_i^{(j)} = 1$ for each $j$.

If all the probabilities $p_i^{(j)} = \textnormal{Tr}[\rho \cdot \Pi_i^{(j)}]$ of observing each POVM outcome were known exactly, the fidelity $F = \textnormal{Tr}[\rho \cdot \ket{\psi}\bra{\psi}]$ could be computed using the decomposition in Eq.~\ref{le_coefficient_condition} according to
\begin{equation}
F = \sum_i\sum_j \alpha_i^{(j)} p_i^{(j)}.
\end{equation}

\noindent This observation motivates the construction of the linear estimator

\begin{equation}
\hat{F} = \sum_i\sum_j \alpha_i^{(j)} \frac{C_i^{(j)}}{n^{(j)}},
\end{equation}
where $n^{(j)}$ is the number of times the $j$th POVM measurement is made during the experiment and $C_i^{(j)}$ is the random variable for the number of times the $i$th outcome is observed during those measurements.

For each $j$, the $C_i^{(j)}$ are distributed multinomially with $n^{(j)}$ trials and probabilities $p_i^{(j)} = \textnormal{Tr}[\rho \Pi_i^{(j)}]$, so the expected value of $\hat{F}$ is
\begin{equation}
\langle \hat{F} \rangle = \sum_i\sum_j \alpha_i^{(j)} \frac{n^{(j)}}{n^{(j)}} \textnormal{Tr}[\rho \cdot \Pi_i^{(j)}] = \textnormal{Tr}[\rho \cdot \ket{\psi}\bra{\psi}],
\end{equation}

\noindent which shows $\hat{F}$ is an unbiased estimator of fidelity. The $j$ runs are independent given $\rho$, so the variance of $\hat{F}$ is

\begin{equation}
\textnormal{Var}(\hat{F}) = \sum_j \sum_{i,i^\prime} \frac{\alpha_i^{(j)} \alpha_{i^\prime}^{(j)}}{(n^{(j)})^2} \textnormal{Cov}(C_i,C_{i^\prime}) = \sum_j \sum_{i,i^\prime} \frac{\alpha_i^{(j)} \alpha_{i^\prime}^{(j)}}{(n^{(j)})^2} n^{(j)}\left(p_i^{(j)}\delta_{ii^\prime} - p_i^{(j)} p_{i^\prime}^{(j)}\right),
\end{equation}

\noindent again using the fact the $C_i$ are multinomially distributed according to probabilities $p_i^{(j)}$.

Because the only constraints on the coefficients $\alpha_i^{(j)}$ are that they satisfy Eq.~\ref{le_coefficient_condition}, an optimal choice of coefficients can be made that minimizes the variance of $\hat{F}$. After choosing to optimize near the reference state $\ket{\psi}\bra{\psi}$ and plugging in $p_i^{(j)} = \textnormal{Tr}[\Pi_i^{(j)} \cdot \ket{\psi}\bra{\psi}]$, the optimal coefficients can be found by solving a quadratic program.

In this experiment, the state space is modeled as a two-qutrit space where each ion has a computational-qubit subspace and a leaked state. The POVM elements are computed numerically assuming particular values of the Poissonian means and depumping and repumping rates calibrated from the reference data. For the purposes of the linear estimator, the fidelity is defined to be $\textnormal{Tr}[\rho \cdot \ket{\psi}\bra{\psi}]$ where all the operators are on the full two-qutrit state space. The initialization-corrected fidelity is computed afterward by assuming a known leakage parameter.

Because of the assumptions about leakage, the constraints on the $\alpha_i^{(j)}$ in Eq.~\ref{le_coefficient_condition} become

\begin{equation}
\sum \alpha_i^{(j)} \Pi_i^{(j)} = \ket{\psi}\bra{\psi} + A \big(\ket{\uparrow a}\bra{\uparrow a} + \ket{a \uparrow}\bra{a \uparrow}\big) + B \big(\ket{\downarrow a}\bra{\downarrow a} + \ket{a \downarrow}\bra{a \downarrow}\big) + C\big(\ket{aa}\bra{aa}\big),
\end{equation}

\noindent for real coefficients $A$,$B$,$C$ that are left as free parameters during the optimization. Now, the associated estimator has an expectation value

\begin{align}
\sum \alpha_i^{(j)} \Pi_i^{(j)} =\,&\textnormal{Tr}[\rho\cdot\ket{\psi}\bra{\psi}] + A \textnormal{Tr}[\rho\cdot (\ket{\uparrow a}\bra{\uparrow a} + \ket{a \uparrow}\bra{a \uparrow})]\\ \nonumber
+ &B \textnormal{Tr}[\rho\cdot (\ket{\downarrow a}\bra{\downarrow a} + \ket{a \downarrow}\bra{a \downarrow})] + C \textnormal{Tr}[\rho\cdot (\ket{aa}\bra{aa})],
\end{align}

\noindent which is shifted from the fidelity by the last three terms. Using the assumptions about leakage and the values of $A$,$B$,$C$ returned by the numerical optimization, those terms can be evaluated and subtracted off to recover an unbiased estimate of fidelity. For example, the symmetric entangled state is assumed to satisfy

\begin{align}
\textnormal{Tr}[\rho\cdot (\ket{\uparrow a}\bra{\uparrow a} + \ket{a \uparrow}\bra{a \uparrow})] &= 2 \epsilon (1-\epsilon),\\
\textnormal{Tr}[\rho\cdot (\ket{\downarrow a}\bra{\downarrow a} + \ket{a \downarrow}\bra{a \downarrow})] &= 0,\\
\textnormal{Tr}[\rho\cdot (\ket{aa}\bra{aa})] &= \epsilon^2,
\end{align}

\noindent where $\epsilon$ is the leakage parameter.

Using this method as a crosscheck, we compare the corrected fidelities we obtain to those reported in the main text using the standard parity analysis method~\cite{Sackett2000} in Table~\ref{table_fidelity} and find good agreement between the methods. We use the same bootstrapping procedure to obtain the 68~\% confidence intervals.

\begin{table}[h]
\begin{center}
 \begin{tabular}{|p{2cm}|p{3cm}|p{1.5cm}|p{3cm}|c|c|} 
 \hline
\textbf{State} & \textbf{Fidelity analysis method} & \textbf{Original data} & \textbf{68~\% confidence interval} & \textbf{Mean} & \textbf{Median} \\ 
 \hline\hline
$\ket{\downarrow\downarrow}+i\ket{\uparrow\uparrow}$ & Parity & 1 & [0.9983, 1] & 1 & 1 \\
\hline
$\ket{\downarrow\downarrow}+i\ket{\uparrow\uparrow}$ & Linear estimator & 1 & [0.9989, 1] & 1 & 1 \\
\hline
$\ket{\downarrow\uparrow}-\ket{\uparrow\downarrow}$ & Parity & 0.9977 & [0.9964, 0.9988] & 0.9976 & 0.9976 \\
\hline
$\ket{\downarrow\uparrow}-\ket{\uparrow\downarrow}$ & Linear estimator & 0.9985 & [0.9968, 1] & 0.9985 & 0.9985 \\
\hline
\end{tabular}
\end{center}
\caption{\label{table_fidelity} Comparison of different fidelity analysis methods. The confidence interval, means, and medians are based on the bootstrap distribution. Values above 1 are truncated.}
\end{table}

\subsection{Fidelity analysis bias estimation\label{sec:bias}}

\begin{figure}
    \centering
    \includegraphics[width=0.5
    \textwidth]{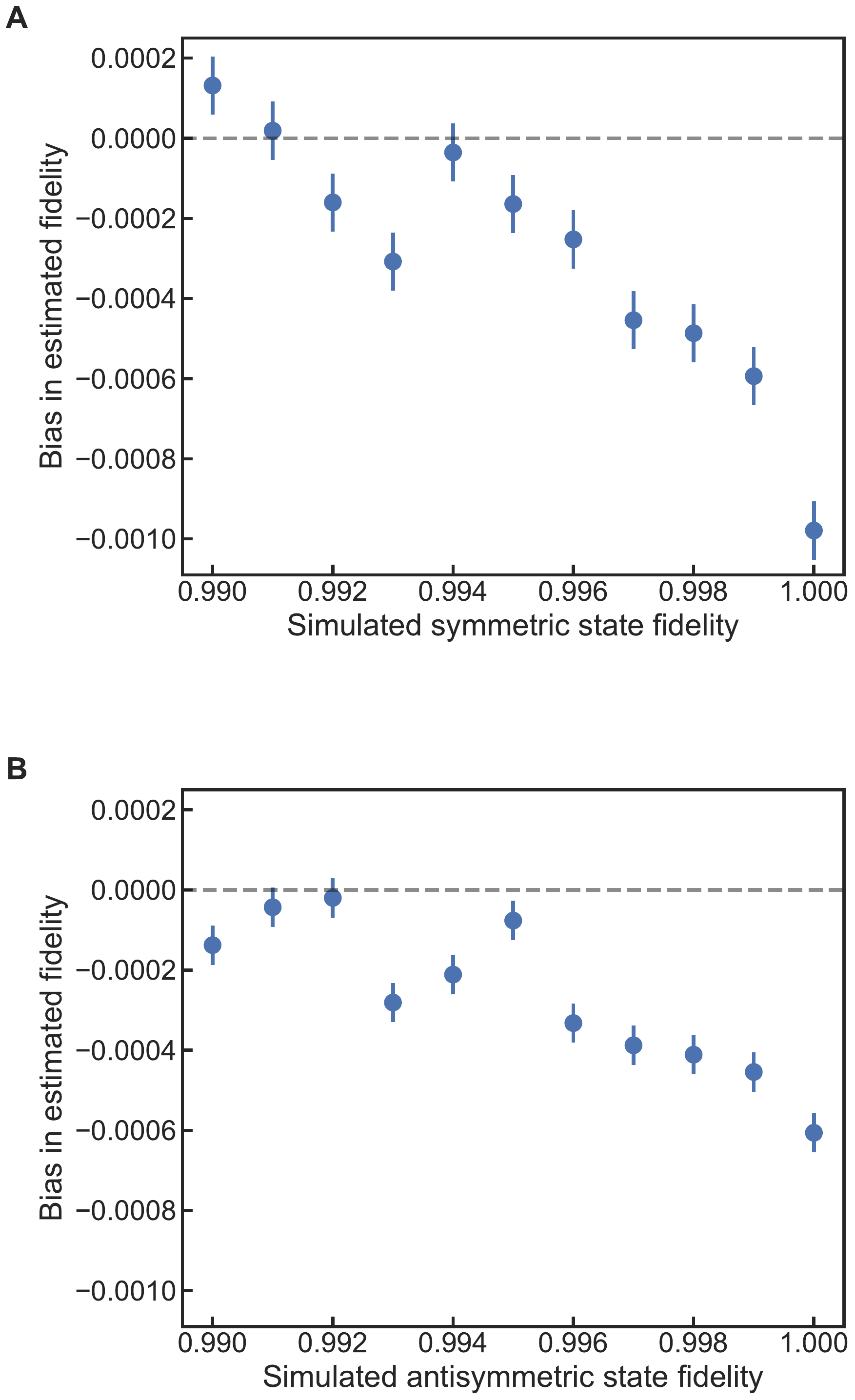}
    \caption{We generate simulated data with identical statistics to the experiment, with different underlying fidelities for both the symmetric (\textbf{A}) and antisymmetric (\textbf{B}) entangled states. We plot the bias in estimated fidelity, the difference between the estimated fidelity and the simulated fidelity, for varying simulated fidelities. A negative bias corresponds to an estimated fidelity that is lower than the simulated fidelity. For each simulated fidelity, we generate 1000 sets of simulated data, including the effects of leakage with similar parameters to what was observed experimentally. For both symmetric and antisymmetric states, the simulated data were generated from a calculated density matrix for the final state at the end of the entangling operation. This density matrix was derived assuming that motional dephasing was the source of entangling gate infidelity. We see an overall negative bias in our estimated fidelity, whose magnitude increases as the underlying fidelity approaches 1.}
    \label{fig_simulated}
\end{figure}

We created simulated data using QuTip\cite{Johansson2013} to validate our fidelity analysis for both the symmetric and antisymmetric entangled states\cite{Srinivas2020}. We varied both the underlying fidelity of the states created due to gate errors as well as leakage errors. We used the simulations to verify that our analysis is either unbiased or underestimates the fidelity of the states we create as shown in Fig.~\ref{fig_simulated}. We find that the parity analysis method is in general more negatively biased (towards a lower fidelity than the true fidelity) compared to the linear estimator. More details of these simulations and their fidelity analysis can be found in Ref.~\cite{Srinivas2020}.

\section{Estimates for leading sources of error}

In this section, we discuss the main sources of error in our entangling operation, which are estimated using QuTip simulations. More details can be found in Ref.~\cite{Srinivas2020}.

\subsection{Motional dephasing}

Frequency fluctuations of the motional frequency $\omega_r$ during the entangling operation will cause dephasing of the ion motion and give rise to an error in the resulting state. This dephasing can usually be measured by implementing a Ramsey experiment on the ion motion. However, this measurement is challenging given our motional frequency drifts and the limited two-ion trap lifetime. We instead use an analysis of the squeezing of the ion motion~\cite{Burd2020} to estimate this dephasing, giving a motional coherence time of about 64\,ms. This dephasing results in an error of $5.8\times10^{-4}$ in the entangled state fidelity.

\subsection{Motional frequency drifts}

The motional frequency $\omega_r$ can drift by several kHz over approximately 7 minutes, the time required to take an entangling operation dataset.  This effect appears to be related to charging of the trap surface from the 280 nm light used for laser cooling.  The motional frequency drift needs to be carefully tracked to achieve the highest fidelities. We perform calibration measurements of the motional frequency every few seconds, interleaved with sets of entangling operation trials, and implement a simple linear feedforward to track and predict the drifts in $\omega_r$. The feedforward prediction, which is used to adjust the frequencies of the qubit control signals for the entangling operation experiment, can be compared after the fact with the record of measured motional frequencies.  The mean of the difference between the prediction and the actual frequency is 3.4\,Hz, with a standard deviation of $\approx50$\,Hz.  These motional frequency fluctuations give rise to an estimated infidelity of approximately $3\times10^{-5}$ in the final Bell state. In our experiment, the feedforward adjusts the value of $\delta$ and keeps $\omega_g$ fixed. This may cause some additional reduction in fidelity at specific $\delta$ values based on their relation to the duration and shape of the rising and falling microwave pulse edges~\cite{Sutherland2019}. 

\subsection{Motional heating}

Finally, we analyze the effect of heating. The motional heating rate of the out-of-phase mode used for the gate is so low that we are unable to measure it precisely; however 1 quanta/s is a conservative upper limit based on measurements. This heating rate will result in an error of $3\times10^{-5}$ in the entangled state fidelity.

\end{document}